\documentclass[12pt]{article}
\usepackage{amsmath,amssymb}
\newcommand{\bsi}{\boldsymbol{\psi}}
\newcommand{\bhi}{\boldsymbol{\varphi}}
\numberwithin{equation}{section}

\title{{\bf Statement of uncertainty principle for quantum
measurements in terms of the R\'{e}nyi entropies}}

\author{Alexey E. Rastegin}
\date{\small Department of Theoretical Physics, Irkutsk State
University, Gagarin Bv. 20, Irkutsk 664003, Russia}

\begin{document}
\maketitle
\begin{abstract}
The aim of the work is to give the explicit proofs of the
R\'{e}nyi--entropy uncertainty relations presented in the previous
work [A. Rastegin, arXiv:0805.1777]. The relations with both the
state-dependent and state-independent entropic bounds are proved.
For a pair of POVM measurements the two relations are obtained. 
The first of them is generalization of the known results, whereas
the second is quite alternative. It is shown that both these relations are
meaty. The important case of POVM's with one-rank elements is extra
discussed. The measurements designed for distinction between two
non-orthogonal quantum states are considered as examples.

\vspace{3mm}
03.65.Ta, 03.67.-a
\end{abstract}

\protect\section{Introduction}

The Heisenberg uncertainty principle \cite{heisenberg} is primary and
most known of those results that emboss the fundamental distinctions
of the quantum world from the classical world. Conceptual development
of quantum theory has lead to a number of related conclusions such as
the Bell inequalities \cite{bell1,bell2}, the quantum Zeno effect
\cite{misra}, the no-cloning theorem \cite{wootters}, the
interaction-free measurement \cite{vaidman,vaidman1}, the no-deleting
principle \cite{pati} and the no-hiding theorem \cite{braun}. The
general quantitative form of the uncertainty principle was given by
Robertson \cite{robert}. Suppose ${\mathsf{A}}$ and ${\mathsf{B}}$
are two observables measured in the quantum state $\bsi$. The 
standard deviations $\Delta{\mathsf{A}}$ and $\Delta{\mathsf{B}}$ of
the two probability distributions then satisfy \cite{robert}
\begin{equation}
\Delta{\mathsf{A}}\>\Delta{\mathsf{B}}\geq
\frac{1}{2}\>|\langle\bsi\>,[{\mathsf{A}}{,}{\mathsf{B}}]\bsi\rangle|
\ , \label{rob}
\end{equation}
where $\langle\cdot{,}\cdot\rangle$ denotes the inner product. This
quantitative formulation of the uncertainty principle is best known
among physicists. Due to a variety of measurement scenarios, many
relations have been stated in more detailed terms
\cite{busch,hilge,hall1}. For example, Bohr's principle of
complementarity \cite{bohr} has been quantified by uncertainty
relations (see papers \cite{horn,kwek,diaz} and references therein).

The entropic uncertainty relations provide an alternative way to
express quantitatively the Heisenberg principle. In many instances
these relations are more useful characterization. The first entropic
relation was proposed by Hirschman \cite{hirs}. Namely, he obtained
the relation for position and momentum in terms of the Shannon
entropies. Hirschman also conjectured an improvement of his result.
This conjecture has been proved by Beckner \cite{beck} and by
Bialynicki-Birula and Mycielski \cite{birula1}. The concrete
calculations of the position and momentum Shannon entropies for the
harmonic oscillator and the hydrogen atom have been made by the
writers of Ref. \cite{yanez}. But important as the case of
position-momentum is, it is not able to give understanding the
limitations on the information characteristics of measurements in all
respects.

In the general context, entropic formulation of the uncertainty
principle was considered by Deutsch \cite{deutsch}. He emphasized
that the right-hand side of Eq. (\ref{rob}) is not a fixed lower
bound but is itself a function of $\bsi$. Indeed, if the state
$\bsi$ is an eigenstate of ${\mathsf{A}}$  then
$\Delta{\mathsf{A}}=0$ and the right-hand side of Eq. (\ref{rob})
vanishes. So no bound on $\Delta{\mathsf{B}}$ is imposed by Eq.
(\ref{rob}) \cite{maass}. Deutsch obtained a state-independent lower
bound on the sum of the Shannon entropies of two probability
distributions generated by measurement of two observables without
degeneracy. Of late years, many other investigations of entropic
uncertainty relations have been made. The list of corresponding
references can be found in the previous work of the present author
\cite{rast}. Here we just mention only several papers of interest.

It turned out that the entropic uncertainty relation given in Ref.
\cite{deutsch} can significantly be improved. The sharpened relation
has been conjectured by Kraus \cite{kraus} and then established by
Maassen and Uffink \cite{maass}. However, the formulation stated in
Ref. \cite{maass} deals with two non-degenerate observables. A
relevant extension to the case of two degenerate observables has been
obtained by Krishna and Parthasarathy \cite{krishna}. Using Naimark's
theorem, they also generalized their entropic uncertainty relation to
the case of a pair of arbitrary measurements. The entropic uncertainty
relations for sets of $N+1$ complementary observables in 
$N$-dimensional Hilbert space were obtained \cite{ruiz93,ruiz95}. In 
more recent paper \cite{massar}, Massar briefly considered the entropic
uncertainty relation in terms of the Shannon entropies for POVM's, whose
elements are all rank one \cite{massar}.

Together with the Shannon entropy, other information entropies are
extensively used in the literature. One of them is the R\'{e}nyi
entropy. In Ref. \cite{larsen}, Larsen derived uncertainty relations in terms
of the so-called purities, which are directly connected with R\'{e}nyi's
entropy of order two. Bialynicki-Birula obtained the uncertainty relations
in terms of R\'{e}nyi's entropies for the position--momentum and
angle--angular momentum pairs \cite{birula3}. In the previous work
\cite{rast}, the present author have posed uncertainty relations for
a pair of arbitrary measurements and for a single measurement in the
form of inequalities using the R\'{e}nyi entropies. The aim of the
given work is to ensure careful proofs of these relations. In
addition, more information on the subject of the entropic uncertainty
relations is provided.

\protect\section{Background}

We shall now describe the notation that is used throughout the text.
By ${\cal{H}}$ we denote finite-dimensional Hilbert space. A state of
quantum system is described by density matrix. Recall that a density
matrix is positive semidefinite matrix with unit trace. Let $\alpha>0$
and $\alpha\not=1$; then the R\'{e}nyi entropy of order $\alpha$ 
of probability distribution $\{p_i\}$ is defined by \cite{renyi}
\begin{equation}
H_{\alpha}(p):=\frac{1}{1-\alpha}\> \ln\left\{
\sum\nolimits_i p_i^{\alpha}
\right\} \ . \label{renent}
\end{equation}
This information measure is a nonincreasing function of order
$\alpha$; that is, if $\alpha<\beta$ then $H_{\alpha}\geq H_{\beta}$
\cite{renyi}. The limit $\alpha\to1$ recovers the Shannon entropy
\begin{equation}
H_1(p):=-\sum\nolimits_i p_i\ln p_i
\ . \label{shanent}
\end{equation}
In the following, orders of R\'{e}nyi's entropies are assumed to be
different from one. The relations for the Shannon entropies can
thereupon be obtained by taking the limit $\alpha\to1$ in the final
inequalities. Unlike the Shannon entropy, the R\'{e}nyi entropy is
not a concave function of the probability distribution. More
precisely, for $\alpha>1$ the R\'{e}nyi entropy $H_{\alpha}(p)$ is
not purely convex nor purely concave \cite{jizba}.

A generalized quantum measurement is described by "Positive
Operator-Valued Measure" (POVM). This is a set $\{{\mathsf{M}}_i\}$
of positive semidefinite matrices satisfying \cite{helstrom,peresq}
\begin{equation}
\sum\nolimits_i {\mathsf{M}}_i={\mathbf{I}}
\ , \label{povmdef}
\end{equation}
where ${\mathbf{I}}$ is the identity matrix. For given measurement
$\{{\mathsf{M}}_i\}$ and quantum state $\rho$, the probability of
$i\,$th outcome is equal to \cite{helstrom,peresq}
\begin{equation}
p_i={\rm tr}\{{\mathsf{M}}_i\rho\}
\ . \label{trac}
\end{equation}
In the mathematical literature, such a set $\{{\mathsf{M}}_i\}$ is
often called "generalized resolution of the identity" for the space
${\cal{H}}$ (for a discussion, see Refs. \cite{holevo,glaz}). In the
particular case of orthogonal projections, one is called "orthogonal
resolution of the identity" \cite{holevo,glaz}.

The R\'{e}nyi entropy $H_{\alpha}({\mathsf{M}}|\rho)$ of generated
probability distribution is then defined by Eqs. (\ref{renent})
and (\ref{trac}). When a quantum state is pure, that is
$\rho=\bsi\>\bsi^{\dagger}$ and $||\bsi||=1$, we will write
$H_{\alpha}({\mathsf{M}}|\bsi)$. In this case,
\begin{equation}
{\rm tr}\{{\mathsf{M}}_i\rho\}=\langle\bsi\>,{\mathsf{M}}_i\bsi\rangle
\ . \label{trac1}
\end{equation}
Let $\{{\mathsf{M}}_i\}$ and $\{{\mathsf{N}}_j\}$ be two POVM's, and
$\bsi$ a pure state. By definition, we put the function
\begin{equation}
f({\mathsf{M}}{,}{\mathsf{N}}|\bsi):=\underset{ij}{\max}\>
||{\mathsf{M}}_i^{1/2}\bsi||^{-1}\>||{\mathsf{N}}_j^{1/2}\bsi||^{-1}
\>|\langle{\mathsf{M}}_i\bsi\>,{\mathsf{N}}_j\bsi\rangle|
\ , \label{fmnpsi}
\end{equation}
where the maximum is taken over those values of labels $i$ and $j$
that satisfy $||{\mathsf{M}}_i^{1/2}\bsi||\not=0$ and
$||{\mathsf{N}}_j^{1/2}\bsi||\not=0$. In the case of mixed state
$\rho$ with the spectral decomposition
\begin{equation}
\rho=\sum\nolimits_{\lambda} \lambda\>\bsi_{\lambda}\bsi_{\lambda}^{\dagger}
\label{rhodec}
\end{equation}
we further define
\begin{equation}
f({\mathsf{M}}{,}{\mathsf{N}}|\rho):=\underset{\lambda}{\max}
\>f({\mathsf{M}}{,}{\mathsf{N}}|\bsi_{\lambda})
\ . \label{fpqrho}
\end{equation}

In order to prove the entropic relation, we shall use Riesz's
theorem. A version of Riesz's theorem is posed as follows
\cite{hardy}. (The below formulation is slightly modified in
comparison with the one given in Ref. \cite{hardy}.) Let
${\mathsf{x}}\in{\mathbb{C}}^n$ be $n$-tuple of complex numbers $x_j$
and let $t_{ij}$ be entries of matrix ${\mathsf{T}}$ of order
$m\times n$. Define $\eta$ to be maximum of the set $|t_{ij}|$, that
is
\begin{equation}
\eta:=\underset{ij}{\max}\>|t_{ij}|
\ . \label{etadef}
\end{equation}
To each ${\mathsf{x}}$ assign $m$-tuple
${\mathsf{y}}\in{\mathbb{C}}^m$ with elements
\begin{equation}
y_i(x):=\sum_{j=1}^{n} t_{ij}\> x_j \quad (i=1,\ldots,m)
\ . \label{lintran}
\end{equation}
So the fixed matrix ${\mathsf{T}}$ describes a linear transformation
${\mathbb{C}}^n\rightarrow{\mathbb{C}}^m$. For any $b\geq1$ we also
define
\begin{equation}
S_b(x):=\left\{ \sum\nolimits_j |x_j|^b
\right\}^{1/b}. \label{sadef}
\end{equation}

{\bf Lemma 1} {\it Suppose the matrix ${\mathsf{T}}$ satisfies}
\begin{equation}
\sum\nolimits_i |y_i|^2\leq\sum\nolimits_j |x_j|^2
\label{suppos}
\end{equation}
{\it for all ${\mathsf{x}}\in{\mathbb{C}}^n$; then}
\begin{equation}
S_a(y)\leq{\eta}^{(2-b)/b}\>S_b(x)
\ , \label{riesz}
\end{equation}
{\it where $1/a+1/b=1$ and $1<b<2$.}

In Ref. \cite{hardy} this result is appeared as theorem 297. Note
that Riesz's theorem has been extended to infinite-dimensional spaces
by Thorin. In functional analysis one is known as the Riesz-Thorin
interpolation theorem \cite{berg}. Of course, the above statement can
be obtained from the Riesz-Thorin theorem. The needed reasons are
contained in Ref. \cite{krishna} (see the proof of theorem 2.1
therein). But the authors of Ref. \cite{krishna} do not formulate the
above statement explicitly as an individual result.

\protect\section{Projective measurements}

A projective measurement is described by "Projector-Valued Measure"
(PVM). This is a set $\{{\mathsf{P}}_i\}$ of Hermitian matrices
satisfying the property
\begin{equation}
{\mathsf{P}}_i{\mathsf{P}}_k=\delta_{ik}{\mathsf{P}}_i
\ , \label{pvmdef1}
\end{equation}
where $\delta_{ik}$ is the Kronecker delta, and the completeness
relation
\begin{equation}
\sum\nolimits_i {\mathsf{P}}_i={\mathbf{I}}
\ . \label{pvmdef2}
\end{equation}
The two PVM's $\{{\mathsf{P}}_i\}$ and $\{{\mathsf{Q}}_j\}$ generate
two probability distributions. Due to the properties of projectors,
the probabilities are rewritten as
\begin{align}
p_i^{(\psi)}&=\langle{\mathsf{P}}_i\bsi,{\mathsf{P}}_i\bsi\rangle
\ , \label{pjp} \\
q_j^{(\psi)}&=\langle{\mathsf{Q}}_j\bsi,{\mathsf{Q}}_j\bsi\rangle
\ , \label{qkp}
\end{align}
The proof of the following statement is given in Appendix A.

{\bf Proposition 2} {\it For two projective measurements
$\{{\mathsf{P}}_i\}$ and $\{{\mathsf{Q}}_j\}$ and pure state
${\rm{\bsi}}\in{\cal{H}}$,}
\begin{equation}
H_{\alpha}({\mathsf{P}}|\bsi)+H_{\beta}({\mathsf{Q}}|\bsi)\geq
-2 \ln f({\mathsf{P}}{,}{\mathsf{Q}}|\bsi)
\ , \label{prop1}
\end{equation}
{\it where orders $\alpha$ and $\beta$ satisfy $1/\alpha+1/\beta=2$.}

As it is mentioned above, the R\'{e}nyi entropy is not generally
concave. Therefore, the lower bound (\ref{prop1}) cannot directly be
extended to the case of mixed state. The Minkowski inequality, which
is very helpful result \cite{hardy}, allows to reach the aim. To
clarify the exposition, we consider the mixed state $\sigma$ with the
spectral decomposition
\begin{equation}
\sigma=\lambda\>\bsi\>\bsi^{\dagger}+(1-\lambda)\ \bhi\>\bhi^{\dagger}
\ , \label{sigdec}
\end{equation}
where $0<\lambda<1$. For the given state $\sigma$, the corresponding
probabilities are then rewritten as
\begin{align}
p_i&={\rm tr}\{{\mathsf{P}}_i\sigma\}=\lambda p_i^{(\psi)}+(1-\lambda)p_i^{(\varphi)}
\ , \label{pjp1} \\
q_j&={\rm tr}\{{\mathsf{Q}}_j\sigma\}=\lambda q_j^{(\psi)}+(1-\lambda)q_j^{(\varphi)}
\ . \label{qkp1}
\end{align}
Here the values $p_i^{(\varphi)}$ and $q_j^{(\varphi)}$ are defined
by substituting $\bhi$ for $\bsi$ into Eqs. (\ref{pjp}) and
(\ref{qkp}) respectively. The argumentation of Appendix A, including
Eqs. (\ref{riezs1}) and (\ref{etadef1}), are valid for both the pure
states $\bsi$ and $\bhi$. So, under the same conditions on $\alpha$
and $\beta$, we can write down
\begin{align}
\lambda S_{\alpha}\{p^{(\psi)}\}&\leq{\eta}^{2(1-\beta)/\beta}\lambda S_{\beta}\{q^{(\psi)}\}
\ , \label{riezs1ps} \\
(1-\lambda)S_{\alpha}\{p^{(\varphi)}\}&\leq{\eta}^{2(1-\beta)/\beta}(1-\lambda)
S_{\beta}\{q^{(\varphi)}\}
\ , \label{riezs1ph}
\end{align}
where $\eta$ is now equal to $f({\mathsf{P}}{,}{\mathsf{Q}}|\sigma)$,
that is the maximum among $f({\mathsf{P}}{,}{\mathsf{Q}}|\bsi)$ and
$f({\mathsf{P}}{,}{\mathsf{Q}}|\bhi)$. On this stage the Minkowski
inequality should be used. By $\alpha>1$ and $\beta<1$, there hold
\begin{align}
&S_{\alpha}\{\lambda p^{(\psi)}+(1-\lambda)p^{(\varphi)}\}\leq
\lambda S_{\alpha}\{p^{(\psi)}\}+(1-\lambda)S_{\alpha}\{p^{(\varphi)}\}
\ , \label{mink1ps} \\
&\lambda S_{\beta}\{q^{(\psi)}\}+(1-\lambda)S_{\beta}\{q^{(\varphi)}\}
\leq S_{\beta}\{\lambda q^{(\psi)}+(1-\lambda)q^{(\varphi)}\}
\ . \label{mink1ph}
\end{align}
Summing Eqs. (\ref{riezs1ps}) and (\ref{riezs1ph}), due to
(\ref{mink1ps}) and (\ref{mink1ph}) we finally get the same relation
(\ref{riezs1}) in which the probabilities are already defined by Eqs.
(\ref{pjp1}) and (\ref{qkp1}). By those transformations that have
lead to Eq. (\ref{prop11}), we obtain an entropic relation
\begin{equation}
H_{\alpha}({\mathsf{P}}|\sigma)+H_{\beta}({\mathsf{Q}}|\sigma)\geq
-2 \ln f({\mathsf{P}}{,}{\mathsf{Q}}|\sigma)
\ . \label{prop02}
\end{equation}
The case of mixed state $\rho$ with the spectral decomposition
(\ref{rhodec}) can be considered in the same manner. Then the
following statement takes place.

{\bf Proposition 3} {\it For two projective measurements
$\{{\mathsf{P}}_i\}$ and $\{{\mathsf{Q}}_j\}$ and any mixed state
$\rho$,}
\begin{equation}
H_{\alpha}({\mathsf{P}}|\rho)+H_{\beta}({\mathsf{Q}}|\rho)\geq
-2 \ln f({\mathsf{P}}{,}{\mathsf{Q}}|\rho)
\ , \label{prop2}
\end{equation}
{\it where orders $\alpha$ and $\beta$ satisfy $1/\alpha+1/\beta=2$.}

\protect\section{One of measurement is generalized}

In this section the above result will be extended to the case when
one of two measurement is described by POVM. Elaborating the ideas
of Ref. \cite{krishna}, we shall use the Naimark extension. All the
necessary details are gathered in Appendix B. The following statement
takes place.

{\bf Proposition 4} {\it Let $\{{\mathsf{M}}_i\}$ be a POVM
measurement, and let $\{{\mathsf{Q}}_j\}$ be a PVM measurement. Then
for any mixed state $\rho$}
\begin{equation}
H_{\alpha}({\mathsf{M}}|\rho)+H_{\beta}({\mathsf{Q}}|\rho)\geq
-2 \ln f({\mathsf{M}}{,}{\mathsf{Q}}|\rho)
\ , \label{prop3}
\end{equation}
{\it where orders $\alpha$ and $\beta$ satisfy $1/\alpha+1/\beta=2$.}

{\bf Proof} Substituting ${\mathsf{M}}_i$ for ${\mathsf{E}}_i$ and
${\mathsf{Q}}_j$ for ${\mathsf{G}}_j$ in the formulas of Appendix B,
we will consider the two measurements $\{\widetilde{\mathsf{M}}_i\}$
and $\{\widetilde{\mathsf{Q}}_j\}$ in the enlarged space
$\widetilde{\cal{H}}$. The measurement $\{\widetilde{\mathsf{M}}_i\}$
is projective due to the Naimark theorem. The measurement
$\{\widetilde{\mathsf{Q}}_j\}$ is projective, because the measurement
$\{{\mathsf{Q}}_j\}$ is projective. By the statement of Proposition
3, we then have
\begin{equation}
H_{\alpha}(\widetilde{\mathsf{M}}|\widetilde\omega)+
H_{\beta}(\widetilde{\mathsf{Q}}|\widetilde\omega)\geq
-2 \ln f(\widetilde{\mathsf{M}}{,}\widetilde{\mathsf{Q}}|\widetilde\omega)
\label{prop31}
\end{equation}
for arbitrary mixed state $\widetilde\omega$ in the enlarged space
$\widetilde{\cal{H}}$. To each density matrix of the form
(\ref{rhodec}) assign the density matrix
\begin{equation}
\widetilde\rho=\sum\nolimits_{\lambda}
 \lambda\>\widetilde\bsi_{\lambda}\widetilde\bsi_{\lambda}^{\dagger}
\ , \label{wrhodec}
\end{equation}
where state vector $\widetilde\bsi_{\lambda}$ is defined by
\begin{equation}
\widetilde\bsi_{\lambda}:=
\begin{bmatrix}
 \bsi_{\lambda} \\
 {\mathbf{0}}
\end{bmatrix}
\ . \label{tilbsilam}
\end{equation}
In the particular case of state $\widetilde\rho$ the relation
(\ref{prop31}) is clearly valid. The $\bsi_{\lambda}$'s form the
orthonormal set in the space ${\cal{H}}$. Hence we obtain
\begin{equation}
\widetilde\bsi_{\lambda}^{\dagger}\widetilde\bsi_{\mu}=
\begin{bmatrix}
 \bsi_{\lambda}^{\dagger} & {\mathbf{0}}
\end{bmatrix}
\begin{bmatrix}
 \bsi_{\mu} \\
 {\mathbf{0}}
\end{bmatrix}
=\bsi_{\lambda}^{\dagger}\bsi_{\mu}=\delta_{\lambda\mu}
\ . \label{worthon}
\end{equation}
So the $\widetilde\bsi_{\lambda}$'s form the (incomplete) orthonormal
set in the space $\widetilde{\cal{H}}$. Due to this fact and the
properties of the trace,
\begin{equation}
{\rm tr}\{\widetilde{\mathsf{M}}_i\widetilde\rho\}=
\sum\nolimits_{\lambda}\lambda\>
\langle\widetilde\bsi_{\lambda}\>,\widetilde{\mathsf{M}}_i\widetilde\bsi_{\lambda}\rangle
=\sum\nolimits_{\lambda}\lambda\>
\langle\bsi_{\lambda}\>,{\mathsf{M}}_i\bsi_{\lambda}\rangle=
{\rm tr}\{{\mathsf{M}}_i\rho\}
\ , \label{wtrip}
\end{equation}
where we use Eq. (\ref{eprobi1}). By a similar argument with Eq.
(\ref{geprobi1}),
\begin{equation}
{\rm tr}\{\widetilde{\mathsf{Q}}_j\widetilde\rho\}=
{\rm tr}\{{\mathsf{Q}}_j\rho\}
\ . \label{wtriq}
\end{equation}
In other words, we have $\widetilde{p_i}=p_i$ and
$\widetilde{q_j}=q_j$ for any state of the form (\ref{wrhodec}).
Therefore, the corresponding R\'{e}nyi entropies are related by
\begin{align}
H_{\alpha}(\widetilde{\mathsf{M}}|\widetilde\rho) &=
H_{\alpha}({\mathsf{M}}|\rho)
\ , \label{walpalp} \\
H_{\beta}(\widetilde{\mathsf{Q}}|\widetilde\rho) &=
H_{\beta}({\mathsf{Q}}|\rho)
\ . \label{wbetbet}
\end{align}
By the relevant substitutions into Eq. (\ref{fegbsi2}) and the
definition (\ref{fpqrho}),
\begin{equation}
f(\widetilde{\mathsf{M}}{,}\widetilde{\mathsf{Q}}|\widetilde\rho)=
f({\mathsf{M}}{,}{\mathsf{Q}}|\rho)
\ . \label{fmqwrho}
\end{equation}
The last three equalities are valid for arbitrary matrix of the form
(\ref{wrhodec}) and, therefore, for arbitrary matrix of the form
(\ref{rhodec}). But the latter is general form of density matrix on
the space ${\cal{H}}$. So from Eq. (\ref{prop31}) we immediately
obtain Eq. (\ref{prop3}). $\blacksquare$

\protect\section{The main results}

In this section the entropic uncertainty relations for general case
will be established. We shall first obtain a lower bound on the sum
of entropies of two POVM measurements $\{{\mathsf{M}}_i\}$ and
$\{{\mathsf{N}}_j\}$. Following the argumentation of the previous
section, let us substitute ${\mathsf{M}}_i$ for ${\mathsf{E}}_i$ and
${\mathsf{N}}_j$ for ${\mathsf{G}}_j$ in the formulas of Appendix B.
So we will consider the two measurements
$\{\widetilde{\mathsf{M}}_i\}$ and $\{\widetilde{\mathsf{N}}_j\}$ in
the enlarged space $\widetilde{\cal{H}}$. The measurement
$\{\widetilde{\mathsf{M}}_i\}$ is now projective due to the Naimark
theorem. In general, the measurement $\{\widetilde{\mathsf{N}}_j\}$
is not projective. By the statement of Proposition 4, for PVM
$\{\widetilde{\mathsf{M}}_i\}$ and POVM
$\{\widetilde{\mathsf{N}}_j\}$ we have
\begin{equation}
H_{\alpha}(\widetilde{\mathsf{M}}|\widetilde\omega)+
H_{\beta}(\widetilde{\mathsf{N}}|\widetilde\omega)\geq
-2 \ln f(\widetilde{\mathsf{M}}{,}\widetilde{\mathsf{N}}|\widetilde\omega)
\ . \label{prop32}
\end{equation}
Here $1/\alpha+1/\beta=2$ and $\widetilde\omega$ denotes an arbitrary
mixed state in the enlarged space $\widetilde{\cal{H}}$. Replacing
$\widetilde{\mathsf{Q}}_j$ with $\widetilde{\mathsf{N}}_j$ in Eqs.
(\ref{wtriq}), (á{\ref{wbetbet}) and (\ref{fmqwrho}), we get
corresponding equalities for the considered case. That is,
\begin{align}
H_{\beta}(\widetilde{\mathsf{N}}|\widetilde\rho) &=
H_{\beta}({\mathsf{N}}|\rho)
\ , \label{wbetbet1} \\
f(\widetilde{\mathsf{M}}{,}\widetilde{\mathsf{N}}|\widetilde\rho) &=
f({\mathsf{M}}{,}{\mathsf{N}}|\rho)
\label{fmqwrho1}
\end{align}
for arbitrary matrix of the form (\ref{wrhodec}). Due to Eqs.
(\ref{walpalp}), (\ref{wbetbet1}) and (\ref{fmqwrho1}), from Eq.
(\ref{prop32}) we immediately obtain the following result.

{\bf Theorem 5} {\it Let $\{{\mathsf{M}}_i\}$ and
$\{{\mathsf{N}}_j\}$ be two POVM measurements. Then for arbitrary
mixed state $\rho$, there holdls}
\begin{equation}
H_{\alpha}({\mathsf{M}}|\rho)+H_{\beta}({\mathsf{N}}|\rho)\geq
-2 \ln f({\mathsf{M}}{,}{\mathsf{N}}|\rho)
\ , \label{theor1}
\end{equation}
{\it where orders $\alpha$ and $\beta$ satisfy $1/\alpha+1/\beta=2$.}

The statement of Theorem 5 is generalization of Theorem 2.5 of Ref.
\cite{krishna} in the following two respects. First, this result
deals with the R\'{e}nyi entropies instead of the Shannon entropies.
Second, it is established for arbitrary mixed state. We shall now obtain
the entropic uncertainty relation for a single POVM presented in Ref.
\cite{rast}. Suppose that $\alpha>1$. For arbitrary state $\rho$ we
then have
\begin{equation}
\sum\nolimits_j p_i^{\alpha}
\leq p_{\rm{max}}^{\alpha-1}\sum\nolimits_j\> p_i \nonumber\\
=p_{\rm{max}}^{\alpha-1}
\ , \label{pref}
\end{equation}
where $p_{\rm{max}}$ is the largest among the probabilities $p_i$ and
the normalization condition is used. Hence due to Eq.
(\ref{renent}) we obtain
\begin{equation}
H_{\alpha}({\mathsf{M}}|\rho)\geq -\ln \phi({\mathsf{M}}|\rho)
\ , \label{haphimr}
\end{equation}
where by definition
\begin{equation}
\phi({\mathsf{M}}|\rho):=\underset{i}{\max}\>
{\rm tr}\{{\mathsf{M}}_i\rho\}\equiv p_{\rm{max}}
\ . \label{phidef}
\end{equation}
Because the R\'{e}nyi entropy is a nonincreasing function of order
$\alpha$, Equation (\ref{haphimr}) remains valid for $\alpha<1$. So
we at once get the needed relation.

{\bf Theorem 6} {\it Let $\{{\mathsf{M}}_i\}$ be a POVM measurement.
Then for arbitrary mixed state $\rho$ and any order $\alpha>0$,}
\begin{equation}
H_{\alpha}({\mathsf{M}}|\rho)\geq
-\ln \phi({\mathsf{M}}|\rho)
\ . \label{theor2}
\end{equation}

The both lower bounds in Eqs. (\ref{theor1}) and (\ref{theor2}) are
dependent on the state in which a quantum system was before
measurement. It is easy to obtain state-independent bounds. Such a
form of entropic bound is usually discussed in the literature. Let us
define the norm of operator ${\mathsf{A}}$ by
\begin{equation}
||{\mathsf{A}}||:=\underset{||\bsi||=1}{\max}
||{\mathsf{A}}\bsi||
\ . \label{normdef}
\end{equation}
As it is shown in Ref. \cite{krishna}, there holds
\begin{equation}
|\langle{\mathsf{M}}_i\bsi\>,{\mathsf{N}}_j\bsi\rangle|\leq
||{\mathsf{M}}_i^{1/2}{\mathsf{N}}_j^{1/2}||
\>||{\mathsf{M}}_i^{1/2}\bsi||\>||{\mathsf{N}}_j^{1/2}\bsi||
\ . \label{minjnorm}
\end{equation}
By definition, we put
\begin{equation}
\bar{f}({\mathsf{M}}{,}{\mathsf{N}})
:=\underset{ij}{\max}\>||{\mathsf{M}}_i^{1/2}{\mathsf{N}}_j^{1/2}||
\ . \label{barfdef}
\end{equation}
It follows from Eqs. (\ref{fmnpsi}), (\ref{fpqrho}) and
(\ref{minjnorm}), that
\begin{equation}
f({\mathsf{M}}{,}{\mathsf{N}}|\rho)\leq\bar{f}({\mathsf{M}}{,}{\mathsf{N}})
\ . \label{fbarf}
\end{equation}
Using Eqs. (\ref{theor1}) and (\ref{fbarf}), we then obtain the
desired bound.

{\bf Corollary 7} {\it Let $\{{\mathsf{M}}_i\}$ and
$\{{\mathsf{N}}_j\}$ be two POVM measurements. For arbitrary mixed
state $\rho$ there holds}
\begin{equation}
H_{\alpha}({\mathsf{M}}|\rho)+H_{\beta}({\mathsf{N}}|\rho)\geq
-2 \ln \bar{f}({\mathsf{M}}{,}{\mathsf{N}})
\ , \label{corol7}
\end{equation}
{\it where orders $\alpha$ and $\beta$ satisfy $1/\alpha+1/\beta=2$.}

In the particular case of Shannon entropies this relation was stated 
in Ref. \cite{krishna}, for one-rank projectors it reduces to
the result given by Maassen and Uffink \cite{maass}. The entropic
relations (\ref{theor1}) and (\ref{corol7}) have been proved under 
the condition $1/\alpha+1/\beta=2$. Suppose now that orders $\alpha$
and $\beta$ are arbitrary. Due to Eq. (\ref{theor2}) we can still 
pose the following uncertainty relation.

{\bf Corollary 8} {\it Let $\{{\mathsf{M}}_i\}$ and
$\{{\mathsf{N}}_j\}$ be two POVM's. For any mixed state $\rho$ and
arbitrary orders $\alpha,\beta\in(0;+\infty)$,}
\begin{equation}
H_{\alpha}({\mathsf{M}}|\rho)+H_{\beta}({\mathsf{N}}|\rho)\geq
-\ln\left[\phi({\mathsf{M}}|\rho)\phi({\mathsf{N}}|\rho)\right]
 \ . \label{corol8}
\end{equation}

Note that there is a natural generalization of Eq. (\ref{corol8}) to
more than two POVM's. Finally, we will obtain a state-independent
bound for a single POVM. Due to the Cauchy-Schwarz inequality and the
definition (\ref{normdef}),
\begin{equation}
|\langle\bsi\>,{\mathsf{M}}_i\bsi\rangle|\leq
||\bsi||\>||{\mathsf{M}}_i\bsi||\leq ||{\mathsf{M}}_i||
\label{imnorm}
\end{equation}
for any normalized state $\bsi$. Let us put the function
\begin{equation}
\bar{\phi}({\mathsf{M}}):=\underset{i}{\max}\>||{\mathsf{M}}_i||
\ . \label{barphid}
\end{equation}
By spectral decomposition of $\rho$, the linearity of the
trace and Eq. (\ref{imnorm}),
\begin{equation}
\phi({\mathsf{M}}|\rho)\leq\bar{\phi}({\mathsf{M}})
\ . \label{phbarph}
\end{equation}
Using Eqs. (\ref{theor2}) and (\ref{phbarph}), we lastly obtain
\begin{equation}
H_{\alpha}({\mathsf{M}}|\rho)\geq
-\ln \bar{\phi}({\mathsf{M}})
\ . \label{corol9}
\end{equation}
All the state-dependent and state-independent bounds on the R\'{e}nyi
entropies proved in this section have first been claimed by the
present author without proofs \cite{rast}. So the above material is
supplementary. In Ref. \cite{rast}, the entropic bounds are
illustrated on the example of distinction between non-orthogonal
quantum states. In the following, we shall continue examination of the
obtained entropic relations.

\protect\section{Discussion}

We shall now consider some features of the entropic uncertainty
relations obtained above. As it is pointed out in Refs. 
\cite{deutsch,maass}, the dependence of the bound in 
Eq. (\ref{rob}) on state $\bsi$ leads to some shortcoming.
In a certain sense, the state-dependent entropic bounds in
Eqs. (\ref{theor1}) and (\ref{theor2}) are free from this defect.
That is, if $\bar{f}({\mathsf{M}}{,}{\mathsf{N}})<1$ then for
each state $\rho$ the bound (\ref{theor1}) is nonzero due to
Eq. (\ref{fbarf}). Further, if $\bar{\phi}({\mathsf{M}})<1$ then for
any $\rho$ the bound (\ref{theor2}) is nonzero due to Eq.
(\ref{phbarph}). This situation takes place if and only if each
of POVM elements has only those eigenvalues that are strictly less
than 1. In physical applications such a property usually implies that
no POVM elements are projectors. This is sufficiently common case. 
In fact, the norm of each projector is equal to 1. Further, due to Eq.
(\ref{povmdef}) the operator $({\mathbf{I}}-{\mathsf{M}_k})$
is positive semidefinite for any fixed $k$. Hence we have
$||{\mathsf{M}_k}||\leq1$, that is no eigenvalues of ${\mathsf{M}_k}$
exceed 1. In addition, both the state-dependent bounds
(\ref{theor1}) and (\ref{theor2}) can be stronger than the
state-independent bounds (\ref{corol7}) and (\ref{corol8})
respectively.  

It must be stressed that inequality (\ref{fbarf}) is always saturated
for two POVM's consisting of elements of rank one only. Due to the
Davies theorem \cite{davies}, such measurements are sufficient to
maximize the mutual information. We shall now prove that for POVM's
with only one-dimensional operators the equality holds in Eq.
(\ref{fbarf}) regardless of state $\rho$. Let us assume that
\begin{align}
{\mathsf{M}_i}&=\mu_i\>{\mathsf{m}_i}\>{\mathsf{m}_i^{\dagger}}
\ , \label{mimui} \\
{\mathsf{N}_j}&=\nu_j\>{\mathsf{n}_j}\>{\mathsf{n}_j^{\dagger}}
\ , \label{njnuj}
\end{align}
for all values of labels $i$ and $j$. (Note that no summation is
taken in Eqs. (\ref{mimui}) and (\ref{njnuj}).) Here $0\leq\mu_i\leq
1$, $0\leq\nu_j\leq 1$, and the vectors ${\mathsf{m}_i}$ and
${\mathsf{n}_j}$ are all normalized. For arbitrary pure state
$\bsi$, we then have
\begin{align}
||{\mathsf{M}}_i^{1/2}\bsi|| &=\mu_i^{1/2}
\>|\langle{\mathsf{m}}_i\>,\bsi\rangle|
\ , \label{sqmimui} \\
||{\mathsf{N}}_j^{1/2}\bsi|| &=\nu_j^{1/2}
\>|\langle{\mathsf{n}}_j\>,\bsi\rangle|
\ . \label{sqnjnuj}
\end{align}
Next, we get
$|\langle{\mathsf{M}}_i\bsi\>,{\mathsf{N}}_j\bsi\rangle|
=\mu_i\nu_j
\>|\langle\bsi\>,{\mathsf{m}}_i\rangle
\langle{\mathsf{m}}_i\>,{\mathsf{n}}_j\rangle
\langle{\mathsf{n}}_j\>,\bsi\rangle|$.
Therefore, for any $\bsi$ there holds
\begin{equation}
||{\mathsf{M}}_i^{1/2}\bsi||^{-1}\>||{\mathsf{N}}_j^{1/2}\bsi||^{-1}
\>|\langle{\mathsf{M}}_i\bsi\>,{\mathsf{N}}_j\bsi\rangle|=
\sqrt{\mu_i\nu_j}\>|\langle{\mathsf{m}}_i\>,{\mathsf{n}}_j\rangle|
\ . \label{sqmunu1}
\end{equation}
The latter is simply equal to
$||{\mathsf{M}}_i^{1/2}{\mathsf{N}}_j^{1/2}||$ accordingly the
properties of operator norm and Eqs. (\ref{mimui}) and (\ref{njnuj}).
Combining this with the definitions (\ref{fpqrho}) and
(\ref{barfdef}) finally gives the claimed equality.

Following Ref. \cite{rast}, we consider the example of
discriminating between pure states $\bsi_1\equiv{\mathbf{e}}_0$ and
$\bsi_2\equiv({\mathbf{e}}_0+{\mathbf{e}}_1)/\sqrt{2}$,
where ${\mathbf{e}}_0$ and ${\mathbf{e}}_1$ are two orthonormal
vectors. This example is very particular case of the quantum
hypothesis testing \cite{auden}. In the Helstrom scheme
\cite{helstrom,holevo}, which is not error-free, the optimal measurement
is described by PVM $\{{\mathsf{N}}_1,{\mathsf{N}}_2\}$ with elements
${\mathsf{N}}_{1}={\mathsf{u}}\>{\mathsf{u}}^{\dagger}$ and
${\mathsf{N}}_{2}={\mathsf{v}}\>{\mathsf{v}}^{\dagger}$, where
\begin{align}
{\mathsf{u}} &\equiv\cos(\pi/8)\>{\mathbf{e}}_0
-\sin(\pi/8)\>{\mathbf{e}}_1 \ , \\
{\mathsf{v}} &\equiv\sin(\pi/8)\>{\mathbf{e}}_0
+\cos(\pi/8)\>{\mathbf{e}}_1 \ .
\end{align}
In the error-free discrimination scheme \cite{ivan,dieks,peres1} the
optimal measurement is described by POVM
$\{{\mathsf{M}}_1,{\mathsf{M}}_2,{\mathsf{M}}_3\}$ with elements
\begin{align}
{\mathsf{M}}_1 &=2^{-1/2}(\sqrt2+1)^{-1}
\>({\mathbf{e}}_0-{\mathbf{e}}_1)\>
({\mathbf{e}}_0-{\mathbf{e}}_1)^{\dagger} \ , \\
{\mathsf{M}}_2 &=\sqrt2\>(\sqrt2+1)^{-1}
\>{\mathbf{e}}_1\>{\mathbf{e}}_1^{\dagger} \ , \\
{\mathsf{M}}_3 &={\mathbf{1}}-{\mathsf{M}}_1-{\mathsf{M}}_2 \ .
\end{align}
The elements of both the above POVM's are all rank one. Due to this
fact and $\bar{f}({\mathsf{M}}{,}{\mathsf{N}})^2=1/2$ \cite{rast},
Theorem 5 gives
\begin{equation}
H_{\alpha}({\mathsf{M}}|\rho)+H_{\beta}({\mathsf{N}}|\rho)\geq \ln 2
\label{exam11}
\end{equation}
for any state $\rho$. Further, $\phi({\mathsf{M}}|\bsi_1)=2^{-1/2}$
and $\phi({\mathsf{N}}|\bsi_1)=2^{-3/2}(\sqrt{2}+1)$ by calculations
\cite{rast}. Corollary 8 then gives
\begin{equation}
H_{\alpha}({\mathsf{M}}|\bsi_1)+H_{\beta}({\mathsf{N}}|\bsi_1)
\geq \ln 4-\ln\,(\sqrt2+1)
\ . \label{exam22}
\end{equation}
The right-hand side of Eq. (\ref{exam11}) is greater than the
right-hand side of Eq. (\ref{exam22}). Thus, for state $\bsi_1$ the
entropic relation (\ref{theor1}) provides more stronger bound than
the entropic relation (\ref{corol8}). The difference between these
bounds is $\ln\,(\sqrt2+1)-\ln 2\approx 0.188$. In turn, the entropic
relation (\ref{corol8}) can be stronger than the entropic relation
(\ref{theor1}). Let us consider the eigenstate $\bhi_3$ of operator
${\mathsf{M}}_3$ which is expressed as
\begin{equation}
\bhi_3=2^{-3/4} \left\{ (\sqrt{2}+1)^{1/2} {\mathbf{e}}_0
+(\sqrt{2}-1)^{1/2} {\mathbf{e}}_1 \right\}
\ . \label{bhi3}
\end{equation}
For this state Theorem 5 poses the same bound given by Eq. 
(\ref{exam11}). By calculations, we further obtain
$\phi({\mathsf{M}}|\bhi_3)=\langle\bhi_3\>,{\mathsf{M}}_3\bhi_3\rangle
=2/(\sqrt{2}+1)$ and 
$\phi({\mathsf{N}}|\bhi_3)=\langle\bhi_3\>,{\mathsf{N}}_1\bhi_3\rangle=
\langle\bhi_3\>,{\mathsf{N}}_2\bhi_3\rangle=1/2$.
Corollary 8 then gives
\begin{equation}
H_{\alpha}({\mathsf{M}}|\bhi_3)+H_{\beta}({\mathsf{N}}|\bhi_3)
\geq \ln\,(\sqrt2+1)
\ . \label{exam33}
\end{equation}
The right-hand side of Eq. (\ref{exam11}) is less than the
right-hand side of Eq. (\ref{exam33}). So, for state $\bhi_3$ the
entropic relation (\ref{corol8}) provides more stronger bound than
the entropic relation (\ref{theor1}). The difference between these
bounds also is $\ln\,(\sqrt2+1)-\ln 2\approx 0.188$. To sum up, we
see that both the entropic uncertainty relations (\ref{theor1}) and
(\ref{corol8}) are independently significant.

Finally, we consider the state-independent bound for a single
POVM. For projective measurement the trivial lower bound on 
the entropy is zero. This bound can exactly be reached. For POVM 
measurement an analogue is ensured by Eq. (\ref{corol9}). Let 
${\mathsf{M}}_0$ be a POVM element such that
$\bar{\phi}({\mathsf{M}})=||{\mathsf{M}}_0||$. It is known that
$H_{\infty}(p)=-\ln p_{\rm{max}}$ \cite{zycz}. Thus, in the case 
$\alpha\gg 1$ the lower bound (\ref{corol9}) is approximately
reached for an eigenstate of ${\mathsf{M}}_0\,$.

\appendix

\protect\section{Proof of Proposition 2}

In this appendix we prove Eq. (\ref{prop1}). Instead of
$p_i^{(\psi)}$ and $q_j^{(\psi)}$, we shall further write $p_i$ and
$q_j$ respectively. For those values of labels $i$ and $j$ that
satisfy $||{\mathsf{P}}_i\bsi||\not=0$ and
$||{\mathsf{Q}}_j\bsi||\not=0$ we define vectors
\begin{align}
{\mathsf{u}}_i &:=||{\mathsf{P}}_i\bsi||^{-1}\>{\mathsf{P}}_i\bsi
\ , \label{uidef} \\
{\mathsf{v}}_j &:=||{\mathsf{Q}}_j\bsi||^{-1}\>{\mathsf{Q}}_j\bsi
\ . \label{vjdef}
\end{align}
With no loss of generality, we can mean that $1\leq i\leq m$ and
$1\leq j\leq n$. The ${\mathsf{u}}_i$'s and the ${\mathsf{v}}_j$'s
form the two orthonormal sets. In general, these sets are not
complete in the space ${\cal{H}}$. By definition, put
\begin{equation}
t_{ij}:=\langle{\mathsf{u}}_i\>,{\mathsf{v}}_j\rangle
\ . \label{tijdef}
\end{equation}
According to Eqs. (\ref{lintran}) and (\ref{tijdef}),
\begin{equation}
y_i(x)=\langle{\mathsf{u}}_i\>,{\mathsf{w}}\rangle
\ , \label{lintran1}
\end{equation}
where ${\mathsf{w}}:=\sum\nolimits_j x_j{\mathsf{v}}_j$ by
definition. It is clear that the vector $\sum\nolimits_i
y_i{\mathsf{u}}_i$ is orthogonal projection of ${\mathsf{w}}$ onto
the subspace spanned by ${\mathsf{u}}_i$'s. So we get
\begin{equation}
\sum\nolimits_i |y_i|^2 \leq ||{\mathsf{w}}||^2
\ . \label{suppos1}
\end{equation}
On the other hand, due to the definition of ${\mathsf{w}}$ and
$\langle{\mathsf{v}}_j\>,{\mathsf{v}}_k\rangle=\delta_{jk}$ we have
\begin{equation}
||{\mathsf{w}}||^2=\sum\nolimits_j  |x_j|^2
\ . \label{suppos2}
\end{equation}
Therefore, the condition (\ref{suppos}) is satisfied for all
${\mathsf{x}}\in{\mathbb{C}}^n$. So we can apply Eq. (\ref{riesz}).
We shall now use this result for the values
\begin{align}
y_i &=||{\mathsf{P}}_i\bsi||
\ , \label{yival} \\
x_j &=||{\mathsf{Q}}_j\bsi||
\ . \label{xjval}
\end{align}
Both the PVM's $\{{\mathsf{P}}_i\}$ and $\{{\mathsf{Q}}_j\}$ satisfy
the completeness relation. Substituting this in the identity
$\bsi={\mathbf{I}}\>\bsi$, in terms of the above values we have
\begin{equation}
\bsi=\sum\nolimits_k y_k{\mathsf{u}}_k=\sum\nolimits_j x_j{\mathsf{v}}_j
\ . \label{valval}
\end{equation}
Combining the relation
$\langle{\mathsf{u}}_i\>,{\mathsf{u}}_k\rangle=\delta_{ik}$ with Eq.
(\ref{valval}), we get
\begin{equation}
y_i=\sum\nolimits_j \langle{\mathsf{u}}_i\>,{\mathsf{v}}_j\rangle\>x_j
\ . \label{valval1}
\end{equation}
Thus, the values given by Eqs. (\ref{yival}) and (\ref{xjval}) are
really connected by Eq. (\ref{lintran}) with the matrix elements
(\ref{tijdef}). Further, $p_i=|y_i|^2$ and $q_j=|x_j|^2$ due to Eqs.
(\ref{pjp}) and (\ref{qkp}). Let us put $a=2\alpha$ and $b=2\beta$.
Squaring Eq. (\ref{riesz}), after substitutions we obtain
\begin{equation}
S_{\alpha}(p)\leq{\eta}^{2(1-\beta)/\beta}S_{\beta}(q)
\ , \label{riezs1}
\end{equation}
where $1/\alpha+1/\beta=2$, $1/2<\beta<1$ and
\begin{equation}
\eta=\underset{ij}{\max}\>|\langle{\mathsf{u}}_i\>,{\mathsf{v}}_j\rangle|
\ . \label{etadef1}
\end{equation}
Using the definition of $H_{\alpha}(p)$ by Eq. (\ref{renent}), we get
\begin{equation}
\ln S_{\alpha}(p)=\frac{1-\alpha}{\alpha} \> H_{\alpha}(p)
\ . \label{salhal}
\end{equation}
In the same way the quantities $S_{\beta}(q)$ and $H_{\beta}(q)$ are
related. It then follows from Eq. (\ref{riezs1}) and
$(1-\beta)/\beta>0$ that
\begin{equation}
\frac{(1-\alpha)\beta}{\alpha(1-\beta)}\>
H_{\alpha}(p)\leq 2\ln\eta+H_{\beta}(q)
\ . \label{riezs2}
\end{equation}
When $1/\alpha+1/\beta=2$ and $\alpha,\beta\not=1$, the multiplier of
$H_{\alpha}(p)$ in Eq. (\ref{riezs2}) is equal to $(-1)$. So we can
rewrite Eq. (\ref{riezs2}) as
\begin{equation}
H_{\alpha}(p)+H_{\beta}(q)\geq -2\ln\eta
\ . \label{prop11}
\end{equation}
For projectors we clearly have ${\mathsf{P}}_i={\mathsf{P}}_i^{1/2}$
and ${\mathsf{Q}}_j={\mathsf{Q}}_j^{1/2}$. So due to Eqs.
(\ref{uidef}) and (\ref{vjdef}) the right-hand side of Eq.
(\ref{etadef1}) is equal to $f({\mathsf{P}}{,}{\mathsf{Q}}|\bsi)$.
This concludes the proof for $\alpha>\beta$. By permutation of the
two PVM's, we recover the case when the order of entropy of PVM
$\{{\mathsf{Q}}_j\}$ is greater than the order of entropy of PVM
$\{{\mathsf{P}}_i\}$.

\protect\section{Naimark's extension and related questions}

We shall now describe the version of Naimark's extension stated in
Ref. \cite{partha}. Let $\{{\mathsf{E}}_i\}$ be a set of positive
semidefinite matrices satisfying
\begin{equation}
\sum\nolimits_i {\mathsf{E}}_i={\mathbf{I}}_{\cal{H}}
\ , \label{fresid}
\end{equation}
where ${\mathbf{I}}_{\cal{H}}$ is the identity operator in the space
${\cal{H}}$. Naimark proved that each generalized resolution of the
identity can be realized as an orthogonal resolution of the identity
for the enlarged space $\widetilde{\cal{H}}$ which contains
${\cal{H}}$ as a subspace \cite{holevo,glaz,naimark}. Let us define
\begin{equation}
\widetilde{\cal{H}}:={\cal{H}}\oplus{\cal{L}}
\ , \label{hhl}
\end{equation}
where ${\cal{L}}$ is a space of needed dimensionality. As it is
interpreted by Partha\-sarathy \cite{partha}, we can build
partitioned matrices of the form
\begin{equation}
\widetilde{\mathsf{E}}_i:=
\begin{bmatrix}
 {\mathsf{E}}_i & {\mathsf{R}}_i \\
 {\mathsf{R}}_i^{\dagger} & {\mathsf{L}}_i
\end{bmatrix}
\ , \label{efl}
\end{equation}
so that the $\widetilde{\mathsf{E}}_i$'s are orthogonal projections
in the enlarged space $\widetilde{\cal{H}}$ and
\begin{equation}
\sum\nolimits_i \widetilde{\mathsf{E}}_i=\widetilde{\mathbf{I}}
\ . \label{eresid}
\end{equation}
Here the matrix $\widetilde{\mathbf{I}}$ represents the identity
operator in the space $\widetilde{\cal{H}}$. In Eq. (\ref{efl}) the
orders of submatrices ${\mathsf{R}}_i$ and ${\mathsf{L}}_i$ should be
clear from the context. An arbitrary vector in the enlarged space is
represented by the column
\begin{equation}
\widetilde{\mathsf{u}}=
\begin{bmatrix}
 {\mathsf{u}} \\
 {\mathsf{z}}
\end{bmatrix}
\label{euz}
\end{equation}
with ${\mathsf{u}}\in{\cal{H}}$ and ${\mathsf{z}}\in{\cal{L}}$. The
entries of this column are components of $\widetilde{\mathsf{u}}$
with respect to the orthonormal basis in $\widetilde{\cal{H}}$ that
is obtained by extension of the initial basis in ${\cal{H}}$.

To each $\bsi\in{\cal{H}}$ assign the vector
$\widetilde\bsi\in\widetilde{\cal{H}}$ defined by
\begin{equation}
\widetilde\bsi:=
\begin{bmatrix}
 \bsi \\
 {\mathbf{0}}
\end{bmatrix}
\ . \label{tilbsi}
\end{equation}
Here and below ${\mathbf{0}}$ denotes the matrix of needed order
consisting of all zeros. Following the rules of block multiplication,
we have
\begin{equation}
\widetilde\bsi^{\dagger}\widetilde{\mathsf{E}}_i\widetilde\bsi=
\begin{bmatrix}
 \bsi^{\dagger} & {\mathbf{0}}
\end{bmatrix}
\begin{bmatrix}
 {\mathsf{E}}_i\bsi \\
 {\mathsf{R}}_i^{\dagger}\bsi
\end{bmatrix}
=\bsi^{\dagger}{\mathsf{E}}_i\bsi
\ . \label{eprobi}
\end{equation}
In other words, the probability of getting outcome $i$, equal to
\begin{equation}
\langle\widetilde\bsi\>,\widetilde{\mathsf{E}}_i\widetilde\bsi\rangle
=\langle\bsi\>,{\mathsf{E}}_i\bsi\rangle
\ , \label{eprobi1}
\end{equation}
is not changed under the made extension.

Let $\{{\mathsf{G}}_j\}$ be another resolution of the identity for
the space ${\cal{H}}$. To each ${\mathsf{G}}_j$ assign the operator
$\widetilde{\mathsf{G}}_j$ acting on the space $\widetilde{\cal{H}}$.
In the matrix representation, we define these operators as follows:
\begin{equation}
\widetilde{\mathsf{G}}_1:=
\begin{bmatrix}
 {\mathsf{G}}_1 & {\mathbf{0}} \\
 {\mathbf{0}} & {\mathbf{I}}_{\cal{L}}
\end{bmatrix}
\ , \quad
\widetilde{\mathsf{G}}_j:=
\begin{bmatrix}
 {\mathsf{G}}_j & {\mathbf{0}} \\
 {\mathbf{0}} & {\mathbf{0}}
\end{bmatrix}
\quad (j\not=1)
\ . \label{gil}
\end{equation}
Here the identity matrix ${\mathbf{I}}_{\cal{L}}$ of corresponding
order describes the action of the identity in the subspace
${\cal{L}}$. Because the ${\mathsf{G}}_j$'s form a resolution of the
identity for the space ${\cal{H}}$, we then have
\begin{equation}
\sum\nolimits_j \widetilde{\mathsf{G}}_j=\widetilde{\mathbf{I}}
\ . \label{geresid}
\end{equation}
Further, for all $\widetilde{\mathsf{u}}\in\widetilde{\cal{H}}$ there
holds
\begin{equation}
\widetilde{\mathsf{u}}^{\dagger}\widetilde{\mathsf{G}}_j\widetilde{\mathsf{u}}
=\begin{bmatrix}
 {\mathsf{u}}^{\dagger} & {\mathsf{z}}^{\dagger}
\end{bmatrix}
\begin{bmatrix}
 {\mathsf{G}}_j{\mathsf{u}} \\
 \delta_{j1}{\mathsf{z}}
\end{bmatrix}
={\mathsf{u}}^{\dagger}{\mathsf{G}}_j{\mathsf{u}}+
\delta_{j1}{\mathsf{z}}^{\dagger}{\mathsf{z}}
\ , \label{gepos}
\end{equation}
So each $\widetilde{\mathsf{G}}_j$ is positive semidefinite due to
the positive semidefiniteness of the ${\mathsf{G}}_j$'s. Therefore,
the set $\{\widetilde{\mathsf{G}}_j\}$ is a resolution of the
identity for the space $\widetilde{\cal{H}}$. In addition, if the
resolution $\{{\mathsf{G}}_j\}$ is orthogonal then the resolution
$\{\widetilde{\mathsf{G}}_j\}$ is also orthogonal. By Eq.
(\ref{gepos}), for any state of the form (\ref{tilbsi}) we have
\begin{equation}
\langle\widetilde\bsi\>,\widetilde{\mathsf{G}}_j\widetilde\bsi\rangle
=\widetilde\bsi^{\dagger}\widetilde{\mathsf{G}}_j\widetilde\bsi=
\bsi^{\dagger}{\mathsf{G}}_j\bsi=
\langle\bsi\>,{\mathsf{G}}_j\bsi\rangle
\ . \label{geprobi1}
\end{equation}
Thus, for second measurement the probability of getting outcome $j$
is also not changed under the made extension. To sum up, we can say
the following. Starting with the two POVM measurements
$\{{\mathsf{E}}_i\}$ and $\{{\mathsf{G}}_j\}$, we have constructed
the two measurements $\{\widetilde{\mathsf{E}}_i\}$ and
$\{\widetilde{\mathsf{G}}_j\}$ in the enlarged space
$\widetilde{\cal{H}}$. But the first measurement
$\{\widetilde{\mathsf{E}}_i\}$ is now projective.

For any positive semidefinite operator there exists a unique positive
square root. Further, each positive semidefinite operator is
Hermitian. Using these facts and the definition of the norm, we get
\begin{equation}
||{\mathsf{E}}_i^{1/2}\bsi||^2=
\langle{\mathsf{E}}_i^{1/2}\bsi\>,{\mathsf{E}}_i^{1/2}\bsi\rangle=
\langle\bsi\>,{\mathsf{E}}_i\bsi\rangle
\ . \label{gnorm}
\end{equation}
Combining this with Eq. (\ref{eprobi1}) finally gives
\begin{equation}
||\widetilde{\mathsf{E}}_i^{1/2}\widetilde\bsi||=
||{\mathsf{E}}_i^{1/2}\bsi||
\label{gnorm1}
\end{equation}
for every state of the form (\ref{tilbsi}). In the same manner, due
to Eq. (\ref{geprobi1}) we obtain
\begin{equation}
||\widetilde{\mathsf{G}}_j^{1/2}\widetilde\bsi||=
||{\mathsf{G}}_j^{1/2}\bsi||
\ . \label{enorm1}
\end{equation}
By matrix calculations, we also have
\begin{equation}
\widetilde\bsi^{\dagger}\widetilde{\mathsf{E}}_i\widetilde{\mathsf{G}}_j
\widetilde\bsi=
\begin{bmatrix}
 \bsi^{\dagger} & {\mathbf{0}}
\end{bmatrix}
\begin{bmatrix}
 {\mathsf{E}}_i & {\mathsf{R}}_i \\
 {\mathsf{R}}_i^{\dagger} & {\mathsf{L}}_i
\end{bmatrix}
\begin{bmatrix}
 {\mathsf{G}}_j\bsi \\
 {\mathbf{0}}
\end{bmatrix}
=\bsi^{\dagger}{\mathsf{E}}_i{\mathsf{G}}_j\bsi
\ . \label{fegbsi}
\end{equation}
By Hermiticity of the POVM elements, in terms of inner products one
gives
\begin{equation}
\langle\>\widetilde{\mathsf{E}}_i\widetilde\bsi\>,
\widetilde{\mathsf{G}}_j\widetilde\bsi\rangle
=\langle\>{\mathsf{E}}_i\bsi\>,{\mathsf{G}}_j\bsi\rangle
\ . \label{fegbsi1}
\end{equation}
Together with Eqs. (\ref{gnorm1}) and (\ref{enorm1}) the last
equality implies
\begin{equation}
f(\widetilde{\mathsf{E}}{,}\widetilde{\mathsf{G}}|\widetilde\bsi)=
f({\mathsf{E}}{,}{\mathsf{G}}|\bsi)
\ . \label{fegbsi2}
\end{equation}
Equation (\ref{fegbsi2}) is valid for arbitrary state vector of the
form (\ref{tilbsi}). Here the one fact should be pointed out. By
calculations,
\begin{equation}
\widetilde\bsi^{\dagger}\widetilde{\mathsf{E}}_i\widetilde{\mathsf{E}}_i
\widetilde\bsi=
\bsi^{\dagger}{\mathsf{E}}_i{\mathsf{E}}_i\bsi+
\bsi^{\dagger}{\mathsf{R}}_i{\mathsf{R}}_i^{\dagger}\bsi
\ . \label{bsieebsi}
\end{equation}
The last equality implies that
$||\widetilde{\mathsf{E}}_i\widetilde\bsi||\not=||{\mathsf{E}}_i\bsi||$
in general. It is for this reason that the square roots of operators
are inserted into the fraction denominator in the right-hand side of
Eq. (\ref{fmnpsi}).

\end{document}